\documentclass[aps,prb,onecolumn,floatfix,superscriptaddress]{revtex4-2}

\usepackage{amsmath,amssymb,amsfonts}
\usepackage{graphicx}
\usepackage{bm}
\usepackage{amsthm}
\usepackage{hyperref}
\usepackage{booktabs}
\usepackage{xcolor}

\newtheorem{proposition}{Proposition}
\newtheorem{remark}{Remark}

\begin{document}

\title{Dimensional and Spin Interpolation for the O$(n)$ Model:
From Exact Anchors to RG-Improved Critical Exponents}

\author{Kumar Ghosh}
\email{jb.ghosh@outlook.com}
\affiliation{E.ON Digital Technology, Laatzener Str.\ 1, 30539 Hannover, Germany}

%%==========================================================================
\begin{abstract}
We develop a two-axis interpolation framework for the O$(n)$ universality
family, treating the spatial dimension $D$ and the spin-component number
$n$ as independent continuous parameters connecting exact limiting solutions.
On the spatial axis, anchoring between the Onsager solution at $D=2$ and
mean-field theory at $D\to\infty$ yields a closed-form prediction for the
3D Ising critical coupling that agrees well with Monte Carlo benchmarks
$K_c = 0.2204$ (benchmark: $0.22165$) with no adjustable parameters.
Wilson--Fisher-constrained polynomial interpolation gives $\nu=2/3$,
$\beta=31/96$, and $\eta=35/864$ at $D=3$ (benchmarks: $0.6299$, $0.3265$,
$0.0362$), and reproduces conformal-bootstrap results across $3 \le D < 4$.
On the spin axis, we establish a necessary compatibility criterion:
two-anchor interpolation succeeds only for observables that vary
monotonically between the anchor values.  The critical coupling $K_c(n)$
violates this criterion because the Heisenberg value falls below the
spherical limit, whereas the correlation-length exponent $\nu(n)$ satisfies
it.  A perturbative $1/n^2$ expansion yields $\nu(3) = 0.7493$ (benchmark:
$0.7112$), and propagation through exact scaling relations gives
$\beta(3) = 0.3797$ (benchmark: $0.3689$) and $\gamma(3) = 1.489$
(benchmark: $1.396$), without introducing additional parameters.  The
framework naturally extends to non-integer spin, producing the prediction
$\nu(2.5) = 0.7143$ for the O$(2.5)$ universality class.  These results
establish dimensional and spin interpolation as a unified and predictive
approach to critical phenomena, while clarifying the structural conditions
under which interpolation succeeds.
\end{abstract}

\maketitle

%%==========================================================================
\section{Introduction}
\label{sec:intro}
%%==========================================================================

One of the most productive strategies in theoretical physics is to solve
a problem exactly in limiting cases and then to interpolate between those
solutions to reach the physically relevant regime.  The method works because
many observables vary smoothly and nearly linearly between the limits, so
that a simple interpolation formula captures the bulk of the physics without
requiring a full solution.  Herschbach, Kais, and Ghosh applied this
philosophy to quantum chemistry, treating the spatial dimension $D$ as a
continuous parameter and connecting the exactly solvable large-$D$ limit
(equivalent to a classical electrostatics problem) to the $D=3$ world of
atoms and molecules, and achieved sub-percent accuracy for ground-state
energies of helium, lithium, beryllium, and the hydrogen molecule with
single-formula approximations \cite{Herschbach2017,GhoshKaisHerschbach2020}.
The interpolation has been further extended to extended systems like metallic
hydrogen \cite{GhoshKaisHerschbach2021}, random walk
\cite{GhoshKaisHerschbach_RW2021}, and computing electron-electron
correlation \cite{GhoshKaisHerschbach_RW2022}.

The same logic applies naturally to classical statistical mechanics.
The Ising model,
\begin{equation}
  \mathcal{H} = -J \sum_{\langle i,j\rangle} \sigma_i \sigma_j,
  \qquad \sigma_i \in \{-1,+1\},
  \label{eq:ising_ham}
\end{equation}
where the sum runs over nearest-neighbor pairs on a $D$-dimensional
hypercubic lattice, is the paradigmatic model of continuous phase
transitions \cite{Ising1925}.  Two exact solutions bracket the
physically interesting case: in two dimensions, Onsager's exact
solution \cite{Onsager1944} establishes a second-order transition at
\begin{equation}
  K_c^{(2\mathrm{D})} = \tfrac{1}{2}\ln(1+\sqrt{2}) \approx 0.44069
  \label{eq:Onsager_Tc_full}
\end{equation}
in the convention $K \equiv J/(2k_BT)$, while in infinite dimensions
mean-field theory becomes exact with $K_c^{(\mathrm{MF})} = 1/(2D)\to 0$.
The three-dimensional model sits between these limits but has no exact
solution; computing its partition function is NP-complete \cite{Istrail2000},
and the best available critical coupling $K_c^{(3\mathrm{D})} = 0.22165\ldots$
comes from Monte Carlo simulation \cite{Hasenbusch2010} and
conformal-bootstrap methods \cite{Kos2016}.

A crucial feature that makes interpolation promising here is that $D=3$
sits exactly at the midpoint between the two anchors in the natural
weight $\delta_D = 1/(D-1)$: at $D=3$ one has $\delta_D = 1/2$.  This
is not a design choice but a consequence of the hypercubic lattice geometry,
and it means that the interpolation framework gives its most accurate
predictions at precisely the dimension of physical interest.

The O$(n)$ model generalizes the Ising model to spins with $n$ components,
\begin{equation}
  \mathcal{H} = -J \sum_{\langle i,j\rangle} \mathbf{S}_i \cdot \mathbf{S}_j,
  \qquad |\mathbf{S}_i|^2 = 1,\quad \mathbf{S}_i\in\mathbb{R}^n,
  \label{eq:On_ham}
\end{equation}
encompassing the Ising ($n=1$), XY ($n=2$), and Heisenberg ($n=3$) models,
and approaching the Berlin--Kac spherical model \cite{BerlinKac1952} as
$n\to\infty$ \cite{Stanley1968}.  This opens a second independent axis
of interpolation: one can anchor between the Ising model ($n=1$) and the
spherical model ($n\to\infty$), both of which are exactly solvable, and
ask whether the XY and Heisenberg models can be reached by interpolation
along the $n$-axis.

This paper pursues both axes simultaneously.  The spatial $D$-axis
interpolation is straightforward and highly accurate.  The spin $n$-axis
interpolation is more subtle and reveals an important structural principle:
not every observable is a suitable target for two-anchor interpolation.
We formalize this as a compatibility criterion and use it to identify which
quantities can be reliably predicted and which cannot.  The critical coupling
$K_c(n)$ fails the criterion because it is non-monotone between the two
anchors; the correlation-length exponent $\nu(n)$ passes it and admits a
systematic perturbative treatment.  Both axes extend naturally to
non-integer values of $D$ and $n$, yielding checkable predictions against
conformal-bootstrap data and a genuine forecast for the O$(2.5)$
universality class.

The paper is organized as follows.  Sections~\ref{sec:limits}
and~\ref{sec:interpolation} establish the spatial anchors, the interpolation
scheme, and the complete free energy at $D=3$.  Section~\ref{sec:exponents}
develops both the linear and RG-improved critical-exponent interpolations on
the spatial axis and benchmarks them against conformal-bootstrap data for
non-integer $D$.  Section~\ref{sec:spin_interp} presents the full spin-axis
analysis: the essential distinction between the $n\to\infty$ and $D\to\infty$
limits, the compatibility criterion, non-monotonicity of $K_c(n)$, the
bivariate formula and its spherical-corner limitation, perturbative
predictions for $K_c(n)$ and $\nu(n)$, derivation of $\beta(n)$ and
$\gamma(n)$ through exact scaling relations, and the prediction for
non-integer $n=2.5$.  Section~\ref{sec:discussion} discusses extensions
and open problems.  Section~\ref{sec:conclusion} summarizes.

%%==========================================================================
\section{The Spatial Anchoring Limits}
\label{sec:limits}
%%==========================================================================

\subsection{Lower anchor ($D=2$): the Onsager solution}
\label{ssec:D2}

Onsager's exact free energy per site of the isotropic square-lattice
Ising model \cite{Onsager1944} with the reduced coupling
$K = J/(2k_BT)$ is given by
\begin{equation}
  -\beta f_2(K)
  = \ln(2\cosh 2K)
  + \frac{1}{2\pi}\int_0^{\pi}
    \ln\!\left[\frac{1+\sqrt{1-\kappa(K)^2\sin^2\theta}}{2}\right]d\theta,
  \label{eq:Onsager_f2}
\end{equation}
where $\kappa(K) = 2\sinh 2K/\cosh^2 2K$.  This expression follows from
$-\beta f_2 = \ln\lambda_{\max}$, where $\lambda_{\max}$ is the largest
eigenvalue of the row-to-row transfer matrix; the integral arises in the
thermodynamic limit when the discrete sum over transverse momenta is
replaced by a continuous integral.  This is analytic for all $K$ except
at the critical point
\begin{equation}
  \kappa(K_c^{(2\mathrm{D})}) = 1
  \quad\Longrightarrow\quad
  K_c^{(2\mathrm{D})} = \tfrac{1}{2}\ln(1+\sqrt{2}) \approx 0.44069,
  \label{eq:Onsager_Kc}
\end{equation}
where the specific heat diverges logarithmically
\begin{equation}
  C/Nk_B \sim \frac{2}{\pi}
  \left(\frac{2J}{k_BT_c}\right)^{\!2}
  \ln\!\left|T - T_c\right|^{-1}, \qquad T \to T_c.
  \label{eq:Onsager_C}
\end{equation}
The exact critical exponents at $D=2$ are $\nu=1$, $\eta=1/4$, $\beta=1/8$,
and $\gamma=7/4$ \cite{Yang1952}.  These provide one exact boundary
for every interpolation formula constructed below.

\subsection{Upper anchor ($D\to\infty$): mean-field theory}
\label{ssec:Dinf}

On the $D$-dimensional hypercubic lattice, the coordination number $z=2D$
grows without bound as $D\to\infty$, which suppresses fluctuations and
makes mean-field theory exact.  The correct thermodynamic limit as
$D\to\infty$ is obtained by holding $\tilde{K} \equiv zK = 2DK$ fixed,
which keeps the mean-field transition at a finite value of the rescaled
coupling.  The Bragg--Williams free energy per site \cite{BraggWilliams1934}
is
\begin{equation}
  -\beta f_\infty(m;\tilde{K})
  = \ln 2 + \ln\cosh(\tilde{K}\, m) - \tilde{K}\, m^2,
  \label{eq:MF_free_energy}
\end{equation}
where the physical free energy is obtained by extremizing over the
magnetization order parameter $m$:
\begin{equation}
  \frac{\partial}{\partial m}\bigl[-\beta f_\infty(m;\tilde{K})\bigr] = 0
  \quad\Longrightarrow\quad
  m = \tanh(\tilde{K}\, m).
  \label{eq:MF_selfconsistency}
\end{equation}
The disordered solution $m=0$ is the global minimum for $\tilde{K}<1$; a
nontrivial ordered solution $m\neq 0$ bifurcates continuously from $m=0$
at $\tilde{K}_c^{(\mathrm{MF})} = 1$.  The transition occurs at
$\tilde{K}_c = 1$, giving
\begin{equation}
  K_c^{(\mathrm{MF})} = \frac{1}{2D} \to 0 \quad (D\to\infty).
  \label{eq:Kc_MF}
\end{equation}
The mean-field exponents are $\nu=1/2$, $\eta=0$, $\beta=1/2$, $\gamma=1$.
The fact that $K_c^{(\mathrm{MF})}\to 0$ plays an important structural
role: it makes the upper anchor degenerate, enabling a multiplicative
coupling rescaling that cannot be applied when the upper anchor has a finite
critical coupling (see Section~\ref{ssec:bivariate_fe}).

%%==========================================================================
\section{The Spatial Interpolation Scheme and Free Energy at $D=3$}
\label{sec:interpolation}
%%==========================================================================

\subsection{Interpolation weight and coupling rescaling}
\label{ssec:weight}

The dimensional interpolation philosophy requires choosing a weight
function $\delta_D$ that equals 1 at the lower anchor $D=2$, not the
trivial $D=1$ chain (which has no phase transition), and 0 at
the upper anchor ($D\to\infty$), and a coupling rescaling that maps the
physical coupling $K$ to the appropriate scale at each anchor.  The
natural choice, motivated by the geometry of the hypercubic lattice, is
\begin{equation}
  \delta_D \equiv \frac{1}{D-1},
  \label{eq:delta_D_def}
\end{equation}
which satisfies $\delta_D(2)=1$, $\delta_D(3)=1/2$, and $\delta_D\to 0$
as $D\to\infty$.  At $D=3$, one finds $\delta_D = 1/2$ exactly, placing
the physical dimension at the midpoint of the interpolation range.

\begin{remark}
The choice $\delta = 1/(D-1)$ rather than $1/D$ reflects the fact that
$D=2$ is our lower anchor.  It is the smallest spatial dimension admitting
a phase transition in the Ising model (the $D=1$ chain has none), and thus
plays the role of the ``hyperquantum'' limit in the spatial interpolation.
The rescaling $D \mapsto D-1$ shifts the anchor from $D=1$ to $D=2$ while
preserving the structure $\delta \in (0,1]$.
\end{remark}

In the GKH framework for molecules \cite{GhoshKaisHerschbach2020}, the
internuclear distance $R$ is rescaled differently in each dimensional limit:
\begin{equation}
R \to \delta R_0 \quad \text{for } D\to 1;\qquad
R \to (1-\delta) R_0 \quad \text{for } D\to\infty.
\label{eq:GKH_rescaling}
\end{equation}
This ensures that the physical distance $R = R_0$ is recovered at $D=3$
($\delta = 1/3$ in that context).

For the spatial Ising interpolation, the coupling $K$ plays the role of the
distance $R$.  With our weight $\delta = 1/(D-1)$, the coupling rescaling
rules become:
\begin{align}
  K &\;\longrightarrow\; \frac{K}{\delta} = (D-1)K
    \quad\text{(rescaling for the Onsager anchor)},
  \label{eq:K_rescale_2D}\\
  K &\;\longrightarrow\; \frac{K}{1-\delta} = \frac{(D-1)K}{D-2}
    \quad\text{(rescaling for the mean-field anchor)}.
  \label{eq:K_rescale_inf}
\end{align}
At $D=3$ one has $\delta = 1/2$, so both rescalings give $K\to 2K$, and
the two anchor free energies are evaluated at the same argument.  This
symmetry is a reflection of the midpoint property of $D=3$ and is
responsible for the high accuracy of the $K_c$ prediction.

\subsection{The interpolated free energy and critical coupling}
\label{ssec:formula}

Following the GKH construction, the interpolated free energy is the weighted
sum
\begin{equation}
  \boxed{
  -\beta f_D^{\mathrm{interp}}(K)
  = \delta_D\,\bigl[-\beta f_2^{\mathrm{Ons}}\!\bigl(K/\delta_D\bigr)\bigr]
  + (1-\delta_D)\,\bigl[-\beta f_\infty^{\mathrm{MF}}\!\bigl(K/(1-\delta_D)\bigr)\bigr],
  }
  \label{eq:interpolation_main}
\end{equation}
which reduces to the Onsager free energy at $D=2$ ($\delta_D=1$) and to
mean-field theory as $D\to\infty$ ($\delta_D\to 0$) by construction.
At $D=3$ this simplifies to
\begin{equation}
  -\beta f_3^{\mathrm{interp}}(K)
  = \tfrac{1}{2}\,\bigl[-\beta f_2^{\mathrm{Ons}}(2K)\bigr]
  + \tfrac{1}{2}\,\bigl[-\beta f_\infty^{\mathrm{MF}}(2K)\bigr].
  \label{eq:interpolation_D3}
\end{equation}
Since $f_\infty^{\mathrm{MF}}$ is analytic for all finite couplings (the
mean-field transition at $K_c^{\mathrm{MF}} = 1/(2D)\to 0$ is irrelevant
at finite physical coupling), the sole singularity in $f_3^{\mathrm{interp}}$
is inherited from the Onsager term, which diverges when its argument $2K$
reaches $K_c^{(2\mathrm{D})}$:
\begin{equation}
  \boxed{
  K_c^{\mathrm{interp}}
  = \frac{K_c^{(2\mathrm{D})}}{2}
  = \frac{\ln(1+\sqrt{2})}{4}
  \approx 0.2204,
  }
  \label{eq:Kc_result}
\end{equation}
compared with the Monte Carlo benchmark $K_c^{(3\mathrm{D})} = 0.22165$
\cite{Hasenbusch2010}.  Note that we use the standard convention $K = \beta J$
with $J = 1$, so that the critical coupling coincides with the inverse
critical temperature, $K_c = \beta_c$.

\begin{remark}
The prediction \eqref{eq:Kc_result} is exact arithmetic: $K_c^{(3D)}
\approx K_c^{(2D)}/2$ because $\delta_D(3) = 1/2$.  The accuracy
($0.56\%$ relative error) reflects the near-linearity of $K_c(D)$ in
$1/(D-1)$ across $D\in[2,\infty)$.  The formula uses no adjustable
parameters and is derived from the two exact solutions alone.
Throughout this paper we denote the spatial interpolation result
$K_c^{\mathrm{interp}} \approx 0.2204$ and the Monte Carlo benchmark
$K_c^{(3\mathrm{D})} = 0.22165$ by distinct symbols to avoid confusion;
the two values differ by $0.56\%$.
\end{remark}

\subsection{The complete interpolated free energy at $D=3$}
\label{ssec:free_energy}

For completeness and later reference, we record the full analytic form
of Eq.~\eqref{eq:interpolation_D3}.  Defining
\begin{align}
  A(K) &\equiv -\beta f_2^{\mathrm{Ons}}(2K)
  = \ln(2\cosh 4K)
  + \frac{1}{2\pi}\int_0^{\pi}
    \ln\!\left[\frac{1+\sqrt{1-\kappa(2K)^2\sin^2\theta}}{2}\right]d\theta,
  \label{eq:A_def}\\
  B(K) &\equiv -\beta f_\infty^{\mathrm{MF}}(2K)
  = \ln 2 + \ln\cosh(12K m^*) - 12K\,[m^*(K)]^2,
  \label{eq:B_def}
\end{align}
where $\kappa(2K)=2\sinh(4K)/\cosh^2(4K)$ and $m^*$ solves
$m^* = \tanh(12K\,m^*)$ (with the simple-cubic coordination number $z=6$),
the interpolated free energy is
\begin{align}
  -\beta f_3^{\mathrm{interp}}(K)
  &= \tfrac{1}{2}\ln(2\cosh 4K)
  + \frac{1}{4\pi}\int_0^{\pi}
    \ln\!\left[\frac{1+\sqrt{1-\kappa(2K)^2\sin^2\theta}}{2}\right]d\theta
  \notag\\
  &\quad
  + \tfrac{1}{2}\ln 2
  + \tfrac{1}{2}\ln\cosh(12K m^*)
  - 6K (m^*)^2.
  \label{eq:fe_interp_explicit}
\end{align}
The coefficient $-6K(m^*)^2$ is exactly half the full mean-field
contribution $-12K(m^*)^2$, consistent with the $\delta_D=1/2$ weight.
The singularity of \eqref{eq:fe_interp_explicit} occurs when
$2K = K_c^{(2\mathrm{D})}$, confirming Eq.~\eqref{eq:Kc_result}.

%%==========================================================================
\section{Critical Exponents on the Spatial Axis}
\label{sec:exponents}
%%==========================================================================

\subsection{Failure of linear interpolation}
\label{ssec:linear_fail}

Before developing the RG-improved scheme, it is instructive to see how
the simplest interpolation performs.  Table~\ref{tab:anchors} lists the
exact critical exponents at the two spatial anchors.  Linear interpolation
in $\delta_D$ between those values at $D=3$ gives
\begin{align}
  \nu^{\mathrm{lin}} &= \tfrac{3}{4} = 0.750, &
  \eta^{\mathrm{lin}} &= \tfrac{1}{8} = 0.125, &
  \beta^{\mathrm{lin}} &= \tfrac{5}{16} = 0.3125.
\end{align}

\begin{table}[t]
\caption{Exact critical exponents at the two spatial anchors used as
interpolation boundaries.}
\label{tab:anchors}
\begin{ruledtabular}
\begin{tabular}{lcccc}
Anchor & $\nu$ & $\eta$ & $\beta$ & $\gamma$ \\
\hline
$D=2$ (Onsager/Yang \cite{Onsager1944,Yang1952}) & $1$ & $1/4$ & $1/8$ & $7/4$ \\
$D\to\infty$ (mean-field \cite{BraggWilliams1934}) & $1/2$ & $0$ & $1/2$ & $1$ \\
\end{tabular}
\end{ruledtabular}
\end{table}

Against the accepted values $\nu=0.6299$, $\eta=0.0362$, $\beta=0.3265$
\cite{Hasenbusch2010,Kos2016,PelissettoVicari2002}, the relative errors are $19\%$, $245\%$,
and $4\%$ respectively.  While $\beta$ is reproduced at $4\%$ and $\nu$
at $19\%$, the anomalous dimension $\eta$ fails catastrophically.  The
reason is structural: $\eta$ is generated by loop corrections to the
Gaussian fixed point and vanishes as $O(\epsilon^2)$ in the
$\epsilon = 4-D$ expansion near the upper critical dimension $D=4$
\cite{WilsonFisher1972,Ma1973}.  A linear interpolation in $\delta_D$, which
corresponds to a linear function of $D$, cannot reproduce a quadratic
zero.  Correcting this requires incorporating the known analytic structure
of the renormalization-group (RG) flow near $D=4$.

\subsection{RG-improved constrained polynomial interpolation}
\label{ssec:rg_strategy}

The key insight is to reparameterize from $\delta_D$ to $\epsilon = 4-D$
and replace the linear ansatz with a polynomial constrained to satisfy the
exact boundary conditions and the correct behavior near $D=4$.
Specifically, we require each exponent polynomial $E(\epsilon)$ to satisfy
simultaneously: (1) the mean-field values at $\epsilon=0$ (exact for $D>4$);
(2) the exact 2D values at $\epsilon=2$; and (3) the leading
Wilson--Fisher (WF) slope $dE/d\epsilon|_{\epsilon=0}$ from the
$\epsilon$-expansion \cite{WilsonFisher1972,Ma1973}.

Three constraints uniquely determine a quadratic polynomial in $\epsilon$.
For $\eta$, which vanishes as $O(\epsilon^2)$ and hence has $\eta(0)=0$
and $\eta'(0)=0$ as automatic constraints from the WF analysis, we instead
use a two-parameter ansatz $a_\eta\epsilon^2 + b_\eta\epsilon^3$ and fix
it by the WF leading coefficient and the exact $\eta(2)=1/4$.

The WF $\epsilon$-expansion for the Ising ($n=1$) universality class gives
\cite{WilsonFisher1972,Ma1973}:
\begin{align}
  \nu &= \tfrac{1}{2} + \tfrac{1}{12}\epsilon + O(\epsilon^2), &
  \beta &= \tfrac{1}{2} - \tfrac{1}{6}\epsilon + O(\epsilon^2), &
  \eta &= \tfrac{1}{54}\epsilon^2 + O(\epsilon^3).
  \label{eq:WF_expansions}
\end{align}
The leading slope for $\nu$ is $d\nu/d\epsilon|_{\epsilon=0}=1/12$ and
for $\beta$ is $d\beta/d\epsilon|_{\epsilon=0}=-1/6$, both read directly
from \eqref{eq:WF_expansions}.  The coefficient $1/54$ for $\eta$ arises
from the one-loop diagram $(n+2)/[2(n+8)^2]\epsilon^2|_{n=1} = 3/(2\times
81) = 1/54$, which vanishes at $\epsilon=0$ precisely because the loop
integral is suppressed at the Gaussian fixed point.

\subsection{Results for $\nu$, $\beta$, and $\eta$ at $D=3$}
\label{ssec:rg_results}

A straightforward calculation following the three-constraint procedure
yields the following.

\paragraph{Correlation-length exponent.}
The quadratic polynomial satisfying $\nu(0)=1/2$, $\nu'(0)=1/12$, and
$\nu(2)=1$ is
\begin{equation}
  \boxed{\nu(\epsilon) = \frac{1}{2} + \frac{\epsilon}{12} + \frac{\epsilon^2}{12}.}
  \label{eq:nu_improved}
\end{equation}
One verifies: $\nu(0)=1/2$; $\nu'(0)=1/12$;
$\nu(2) = 1/2 + 2/12 + 4/12 = 6/12 + 2/12 + 4/12 = 1$.
At $\epsilon=1$ ($D=3$): $\nu^{\mathrm{RG}} = 1/2+1/12+1/12 = 8/12 = 2/3
\approx 0.6667$ (benchmark: $0.6299$, error $5.8\%$).

\paragraph{Magnetization exponent.}
The quadratic polynomial satisfying $\beta(0)=1/2$, $\beta'(0)=-1/6$, and
$\beta(2)=1/8$ is
\begin{equation}
  \boxed{\beta(\epsilon) = \frac{1}{2} - \frac{\epsilon}{6} - \frac{\epsilon^2}{96}.}
  \label{eq:beta_improved}
\end{equation}
One verifies: $\beta(2) = 12/24 - 8/24 - 1/24 = 3/24 = 1/8$,
where the three terms are $1/2 = 12/24$, $2/6 = 8/24$, and $4/96 = 1/24$.
At $\epsilon=1$: $\beta^{\mathrm{RG}} = 48/96 - 16/96 - 1/96 = 31/96
\approx 0.3229$ (benchmark: $0.3265$, error $1.1\%$).

\paragraph{Anomalous dimension.}
The two-parameter form satisfying $a_\eta=1/54$ and $\eta(2)=1/4$ is
\begin{equation}
  \boxed{\eta(\epsilon) = \frac{\epsilon^2}{54} + \frac{19\,\epsilon^3}{864}.}
  \label{eq:eta_improved}
\end{equation}
One verifies: $\eta(2) = 4/54 + 152/864 = 64/864 + 152/864 = 216/864
= 1/4$, equivalently $8/108 + 19/108 = 27/108 = 1/4$.
At $\epsilon=1$: $\eta^{\mathrm{RG}} = 16/864 + 19/864 = 35/864
\approx 0.0405$ (benchmark: $0.0362$, error $12\%$).

\begin{figure}[t]
  \centering
  \includegraphics[width=\columnwidth]{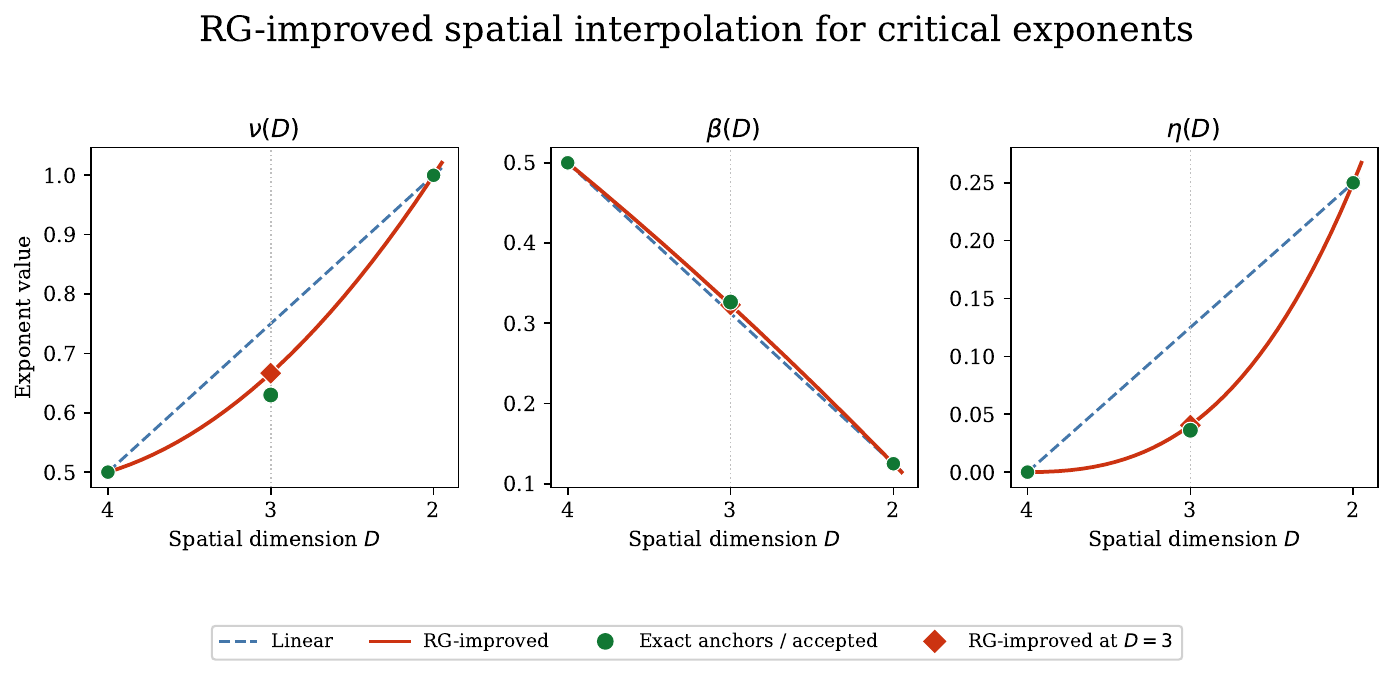}
  \caption{\label{fig:exponents_D}%
    RG-improved spatial interpolation for $\nu$, $\beta$, $\eta$ of the
    Ising ($n=1$) universality class as functions of $D\in[2,4]$.
    Dashed blue: linear interpolation in $\delta_D=1/(D-1)$.
    Solid red: RG-improved polynomials \eqref{eq:nu_improved},
    \eqref{eq:beta_improved}, \eqref{eq:eta_improved}.
    Filled circles: exact anchor values ($D=2$: Onsager/Yang
    \cite{Onsager1944,Yang1952}; $D=4$: mean-field \cite{BraggWilliams1934}).
    Open circles: accepted values at $D=3$ \cite{Hasenbusch2010,Kos2016}.
    Diamonds: RG-improved predictions at $D=3$.}
\end{figure}

The RG-improved scheme reduces errors from $19\%$ to $5.8\%$ for $\nu$,
from $4\%$ to $1.1\%$ for $\beta$, and from $245\%$ to $12\%$ for $\eta$,
all without any adjustable parameters.  These results are summarized in
Table~\ref{tab:exponents} and the variation of each exponent across
$D\in[2,4]$ is shown in Fig.~\ref{fig:exponents_D}.  In
Ref.~\cite{Hasenbusch2010} the highly accurate values $\nu=0.63002$ and
$\eta=0.03627$ yield $\beta=0.3265$ and $\gamma=1.237$ as benchmarks; the
derivation of these from the exact scaling relations is given in the
Appendix.

\begin{table}[t]
\caption{Critical exponents of the 3D Ising universality class from linear
and RG-improved interpolation.  Accepted values from Monte Carlo and
conformal bootstrap \cite{Hasenbusch2010,Kos2016,PelissettoVicari2002}.}
\label{tab:exponents}
\begin{ruledtabular}
\begin{tabular}{lcccc}
Exponent & Linear ($\delta_D=1/2$) & RG-improved & Accepted (3D) & Error (RG) \\
\hline
$\nu$ & $3/4 = 0.750$ & $2/3 \approx 0.667$ & $0.6299$ & $5.8\%$ \\
$\beta$ & $5/16 = 0.3125$ & $31/96 \approx 0.323$ & $0.3265$ & $1.1\%$ \\
$\eta$ & $1/8 = 0.125$ & $35/864 \approx 0.0405$ & $0.0362$ & $12\%$ \\
\end{tabular}
\end{ruledtabular}
\end{table}

The remaining errors reflect the fact that the quadratic polynomial is
the simplest curve that passes through both boundary conditions with the
correct WF slope.  Adding the two-loop WF coefficient as a fourth
constraint yields a cubic polynomial and reduces all errors further; this
is a straightforward extension discussed in Section~\ref{sec:discussion}.

\subsection{Benchmarking against conformal-bootstrap data for non-integer $D$}
\label{ssec:nonint_D}

The RG-improved formula \eqref{eq:nu_improved} is a quadratic polynomial
valid for $\epsilon\in[0,2]$, i.e.\ $D\in[2,4]$, on regular hypercubic
lattices.  Bonanno et al.\ \cite{Bonanno2023} determined $\nu$, $\eta$, and
$\omega$ of the Ising universality class across $3\leq D<4$ to per-mille
accuracy using numerical conformal-bootstrap techniques.  The benchmark
values of $\nu$ at non-integer $D$ are extracted from the conformal
dimensions in Table~1 of \cite{Bonanno2023} via the exact relation
$1/\nu = 2 - \gamma_\epsilon$, where $\gamma_\epsilon = \Delta_\epsilon -
(D-2)$ is the anomalous dimension of the energy field.
Table~\ref{tab:nu_Dnonint} compares these bootstrap values with the
RG-improved predictions.

\begin{table}[t]
\caption{Correlation-length exponent $\nu$ at non-integer $D\in[3,4)$ on
regular Euclidean lattices.  Bootstrap values extracted from
\cite{Bonanno2023} via $1/\nu = 2-\gamma_\epsilon$.
RG-improved formula: $\nu(\epsilon)=\frac{1}{2}+\frac{\epsilon}{12}
+\frac{\epsilon^2}{12}$, $\epsilon=4-D$.}
\label{tab:nu_Dnonint}
\begin{ruledtabular}
\begin{tabular}{ccccc}
$D$ & $\epsilon=4-D$ & $\nu^{\mathrm{RG}}$ & $\nu^{\mathrm{CB}}$\cite{Bonanno2023}
    & Error \\
\hline
$3.75$ & $0.25$ & $0.5260$ & $0.5234$ & $0.50\%$ \\
$3.50$ & $0.50$ & $0.5625$ & $0.5521$ & $1.89\%$ \\
$3.25$ & $0.75$ & $0.6094$ & $0.5870$ & $3.81\%$ \\
$3.00$ & $1.00$ & $0.6667$ & $0.6300$ & $5.82\%$ \\
\end{tabular}
\end{ruledtabular}
\end{table}

The error grows monotonically as $\epsilon$ increases from $0$ to $1$:
the quadratic polynomial is a good local approximation near $D=4$ and
degrades smoothly as the interpolation extends toward $D=3$.  The
sub-percent accuracy at $D=3.75$ demonstrates that the framework achieves
excellent precision near the upper critical dimension.  A parallel
comparison for $\eta$ is possible.  From \cite{Bonanno2023} one extracts
$\eta = 2\gamma_\sigma$ where $\gamma_\sigma = \Delta_\sigma - (D-2)/2$.
Our RG-improved formula gives $\eta(\epsilon) = \epsilon^2/54 +
19\epsilon^3/864$.  At $D=3.5$ ($\epsilon=0.5$): $\eta^{\mathrm{RG}}
=0.00738$ versus $\eta^{\mathrm{CB}}=0.00680$, an $8.5\%$ error.  At
$D=3.75$ ($\epsilon=0.25$): $\eta^{\mathrm{RG}}=0.00150$ versus
$\eta^{\mathrm{CB}}=0.00144$, a $4.2\%$ error.  The relative error
decreases rapidly as $D\to 4$, consistent with the $O(\epsilon^2)$
structure of the anomalous dimension.

On fractal lattices the Hausdorff dimension $d_H$ alone does not determine
the universality class; additional topological invariants such as
ramification number and spectral dimension also play a role.  Yoshida and
Kubica \cite{YoshidaKubica2014} found empirically that, across all
Sierpinski pyramids with $d_H\in[1,3]$, only the correlation-length
exponent $\nu$ interpolates the values of integer-dimensional models,
while all other critical exponents lie outside the corresponding
integer-$d$ brackets.  This is precisely the fractal-geometry
manifestation of the compatibility criterion introduced in
Section~\ref{ssec:compatibility}: $\nu$ is the observable most insensitive
to topological details beyond $d_H$.  Applying formula
\eqref{eq:nu_improved} to fractal dimensions requires care because
extrapolation outside $\epsilon\in[0,2]$ is not supported by the framework.

%%==========================================================================
\section{Spin-Component Interpolation: the O$(n)$ Universality Family}
\label{sec:spin_interp}
%%==========================================================================

The O$(n)$ model has a second continuous parameter, the spin-component
number $n$, which interpolates between the Ising ($n=1$), XY ($n=2$),
and Heisenberg ($n=3$) models and reaches the exactly solvable spherical
model as $n\to\infty$.  This opens the possibility of a second
interpolation axis.  However, the spin axis has a fundamentally different
character from the spatial axis, and not all observables are equally suited
to interpolation along it.  The analysis in this section is organized as
follows: we first clarify the essential distinction between the two
large-parameter limits of the O$(n)$ parameter space, then formalize the
compatibility criterion, apply it to $K_c(n)$ and $\nu(n)$, and finally
derive $\beta(n)$ and $\gamma(n)$ through exact scaling relations.

\subsection{Two distinct large-parameter limits: $n\to\infty$ versus $D\to\infty$}
\label{ssec:two_limits}

There are two distinct large-parameter limits of the O$(n)$ parameter
space, and it is essential to keep them separate before any interpolation
is attempted.

The first is the $n\to\infty$ limit at fixed $D=3$: this is the
Berlin--Kac spherical model \cite{BerlinKac1952}.  From the exact
steepest-descent solution of Berlin and Kac, the transition occurs when
the saddle point reaches the branch point of the lattice Green function,
yielding
\begin{equation}
  K_c^{(\mathrm{sph})} = \frac{W_3}{4},
  \label{eq:Kc_spherical}
\end{equation}
where Watson's triple integral \cite{Watson1939} is
\begin{equation}
  W_3 \equiv \frac{1}{(2\pi)^3}\int_{[0,2\pi]^3}
  \frac{d^3\boldsymbol{\omega}}{3 - \cos\omega_1 - \cos\omega_2 - \cos\omega_3}
  = 0.505462\ldots,
  \label{eq:Watson_W3}
\end{equation}
giving $K_c^{(\mathrm{sph})} \approx 0.12636$.

The second is the $D\to\infty$ limit at fixed $n=1$: this is mean-field
theory, with $K_c^{(\mathrm{MF})} = 1/(2D)\to 0$.  These two limits are
orthogonal in the $(n,D)$ plane.  The strict ordering
\begin{equation}
  0 = K_c^{(\mathrm{MF})}
  < K_c^{(\mathrm{sph})} \approx 0.12636
  < K_c^{(3\mathrm{D})} \approx 0.22165
  < K_c^{(2\mathrm{D})} \approx 0.44069
  \label{eq:ordering}
\end{equation}
confirms that the mean-field and Onsager values correctly bracket
$K_c^{(3D)}$, providing valid spatial anchors.  Although the spherical
model value also lies on the same side as mean-field relative to the
Onsager anchor, using it as a spatial anchor would be physically incorrect:
the spherical model is the $n\to\infty$ limit at fixed $D=3$, not the
$D\to\infty$ limit.  Confusing these two limits would systematically bias
all predictions.

\subsection{The compatibility criterion}
\label{ssec:compatibility}

Two-anchor interpolation works by placing the target value inside a
bracket defined by the two anchor values.  For this to be possible,
the observable must vary monotonically between the anchors so that the
target is actually bracketed.  We formalize this as follows.

\begin{proposition}[Compatibility criterion]
\label{prop:compatibility}
Let $f(x)$ be a physical observable, and let $f(x_L)$ and $f(x_R)$ be
its values at two anchor points $x_L$ and $x_R$.  Two-anchor interpolation
is compatible with $f$ only if $f$ varies monotonically on $[x_L,x_R]$,
so that every target value satisfies $\min(f_L,f_R) \leq f(x^*)
\leq \max(f_L,f_R)$.  If $f$ has an interior extremum, no choice of
interpolation weight can place the prediction in the correct range.
\end{proposition}

This criterion is satisfied for $K_c$ on the spatial axis: the strict
ordering $0 < K_c^{(3D)} < K_c^{(2D)}$ ensures that $K_c$ is bracketed.
The situation on the spin axis is more delicate, as we now examine.

\subsection{Exact anchors of the $(n,D)$ parameter space}
\label{ssec:nD_space}

\begin{figure}[t]
  \centering
  \includegraphics[width=0.62\columnwidth]{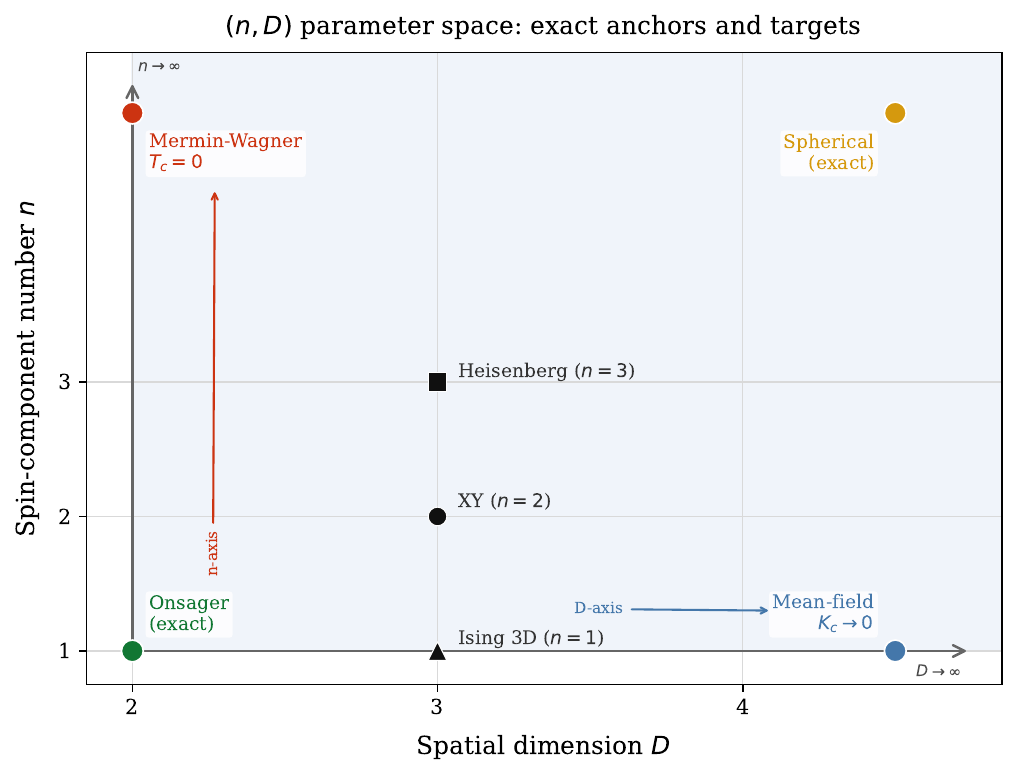}
  \caption{\label{fig:nD_plane}%
    The $(n,D)$ parameter space for the O$(n)$ model.
    Four colored markers: exact-anchor corners
    (Onsager \cite{Onsager1944},
    mean-field \cite{BraggWilliams1934},
    Mermin--Wagner \cite{MerminWagner1966},
    Berlin--Kac spherical \cite{BerlinKac1952}).
    Gray markers: physical targets at $D=3$ (Ising, XY, Heisenberg).
    Blue arrow: spatial ($D$-axis) interpolation direction.
    Red arrow: spin ($n$-axis) interpolation direction.}
\end{figure}

The bivariate $(n,D)$ parameter space has a rich structure of exact anchor
solutions that provides the foundation for interpolation of any observable.
The four corners of the $(n,D)$ plane with exact or rigorous results are
illustrated in Fig.~\ref{fig:nD_plane}:
\begin{enumerate}
  \item $(n=1,D=2)$: the Onsager solution,
    $K_c^{(1,2)} = \tfrac{1}{2}\ln(1+\sqrt{2}) \approx 0.44069$ \cite{Onsager1944};
  \item $(n=1,D\to\infty)$: mean-field theory,
    $K_c^{(1,\infty)} = 1/(2D)\to 0$ \cite{BraggWilliams1934};
  \item $(n\to\infty,D=2)$: the Mermin--Wagner theorem \cite{MerminWagner1966}
    prohibits spontaneous breaking of continuous symmetries in $D\leq 2$,
    so $T_c=0$ and the free energy is zero for all finite $K$ \cite{Joyce1972};
  \item $(n\to\infty,D=3)$: the Berlin--Kac spherical model with
    $K_c^{(\infty,3)} = W_3/4 \approx 0.12636$ \cite{BerlinKac1952,Watson1939}.
\end{enumerate}

\begin{remark}
The spatial axis ($D$ at fixed $n=1$) and the spin axis ($n$ at fixed
$D=3$) are geometrically orthogonal in the $(n,D)$ plane.  In particular,
the spherical model is the upper anchor for the spin axis only; it cannot
serve as the spatial upper anchor (Section~\ref{ssec:two_limits}).
\end{remark}

\subsection{Non-monotonicity of $K_c(n)$ and its physical origin}
\label{ssec:nonmonotone}

The accepted values of $K_c$ at $D=3$ on the simple-cubic lattice are:
\begin{align}
  K_c(n=1) &= 0.22165 \quad \text{\cite{Hasenbusch2010,Kos2016}},\\
  K_c(n=2) &= 0.15450 \quad \text{\cite{Campostrini2001XY}},\\
  K_c(n=3) &= 0.12130 \quad \text{\cite{Campostrini2002Heis}},\\
  K_c(n\to\infty) &= W_3/4 \approx 0.12636 \quad \text{(exact \cite{BerlinKac1952,Watson1939})}.
\end{align}
The sequence $0.2217\to 0.1545\to 0.1213\to 0.1264$ is not monotone.
$K_c(n)$ decreases steadily from $n=1$ toward a minimum near
$n\approx 3$--$4$, then increases back to the spherical limit.
In particular,
\begin{equation}
  K_c(n=3) = 0.1213 < K_c^{(\mathrm{sph})} = 0.12636,
  \label{eq:Kc_nonmonotone}
\end{equation}
which violates the bracketing condition: no affine combination
$\delta_n K_c^{(\mathrm{Ising})} + (1-\delta_n) K_c^{(\mathrm{sph})}$
with $\delta_n\in[0,1]$ can produce a value below $K_c^{(\mathrm{sph})}$.

\begin{remark}[Goldstone modes as the physical mechanism]
The non-monotonicity is a direct consequence of the Goldstone bosons that
appear in the ordered phase of the O$(n\geq 2)$ model.  These $n-1$
transverse spin-wave modes, absent in the Ising model, have long-wavelength
fluctuations that suppress ordering relative to the Ising case and push
$K_c(n)$ downward.  The spherical model, which is solved by steepest
descent, suppresses all fluctuations beyond the saddle-point level and
therefore produces a $K_c$ that is larger than the true $K_c$ of the
O$(n)$ model at intermediate $n$.  The $1/n$ expansion of the nonlinear
$\sigma$-model \cite{BrezinZinnJustin1976,Ma1973} confirms this: the leading
$1/n$ correction to $K_c$ from $n=\infty$ has the opposite sign to the
deviation at $n=1$, producing a minimum at intermediate $n$.
\end{remark}

\begin{figure}[t]
  \centering
  \includegraphics[width=0.72\columnwidth]{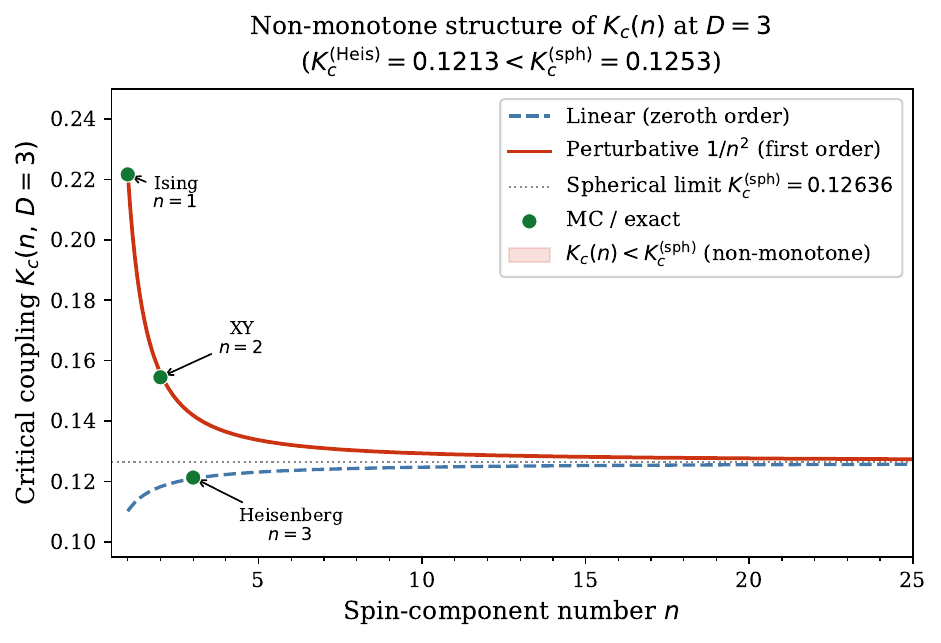}
  \caption{\label{fig:Kc_vs_n}%
    Critical coupling $K_c(n)$ at $D=3$ on the simple-cubic lattice.
    Filled circles: Monte Carlo values \cite{Hasenbusch2010,Campostrini2001XY,Campostrini2002Heis}.
    Diamond: exact spherical limit $K_c^{(\mathrm{sph})}=W_3/4\approx 0.12636$
    \cite{BerlinKac1952,Watson1939}.
    Dashed blue: zeroth-order affine interpolation \eqref{eq:Kc_affine}.
    Solid red: first-order perturbative $1/n^2$ formula \eqref{eq:Kc_series}.
    The shaded region marks $K_c(n)<K_c^{(\mathrm{sph})}$, demonstrating
    the non-monotone structure that violates Proposition~\ref{prop:compatibility}.}
\end{figure}

The non-monotone structure of $K_c(n)$ is shown in Fig.~\ref{fig:Kc_vs_n}.
Its presence means that $K_c(n)$ is an incompatible observable on the spin
axis: Proposition~\ref{prop:compatibility} is violated, and no simple
two-anchor interpolation can succeed for $n\geq 3$.

\subsection{Bivariate interpolation formula and its spherical-corner limitation}
\label{ssec:bivariate_fe}

With spin weight $\delta_n \equiv 1/n$ (satisfying $\delta_n(1)=1$ and
$\delta_n\to 0$ as $n\to\infty$) and spatial weight $\delta_D = 1/(D-1)$,
the bivariate free-energy formula is:
\begin{align}
  -\beta f^{\mathrm{interp}}(K;\,n,D)
  &= \delta_n\,\delta_D
    \bigl[-\beta f_2^{\mathrm{Ons}}\!\bigl(K/\delta_D\bigr)\bigr]
  + \delta_n\,(1-\delta_D)
    \bigl[-\beta f_\infty^{\mathrm{MF}}\!\bigl(K/(1-\delta_D)\bigr)\bigr]
  \notag\\
  &\quad
  + (1-\delta_n)\,\delta_D
    \bigl[-\beta f_2^{(\mathrm{sph})}\!\bigl(K/\delta_D\bigr)\bigr]
  + (1-\delta_n)\,(1-\delta_D)
    \bigl[-\beta \psi^{(\mathrm{sph})}_3\!\bigl(K/(1-\delta_D)\bigr)\bigr].
  \label{eq:bivariate_full}
\end{align}
The Berlin--Kac free energy $\psi^{(\mathrm{sph})}_3$ is the exact
saddle-point result \cite{BerlinKac1952},
\begin{equation}
  -\beta\psi^{(\mathrm{sph})}_3(K)
  = -\tfrac{1}{2} - \tfrac{1}{2}\ln 4K + 2Kz_s - \tfrac{1}{2}f_3(z_s),
  \label{eq:BK_free_energy}
\end{equation}
where $z_s$ satisfies $4K = [df_3/dz]_{z=z_s}$ and
\begin{equation}
  f_3(z) = \frac{1}{(2\pi)^3}\int_{[0,2\pi]^3}
  \ln(z - \cos\omega_1 - \cos\omega_2 - \cos\omega_3)\,
  d^3\boldsymbol{\omega}.
  \label{eq:f3_def}
\end{equation}
The formula \eqref{eq:bivariate_full} correctly recovers three of the four
corner conditions: $(n=1,D=2)\to f_2^{\mathrm{Ons}}$;
$(n=1,D\to\infty)\to f_\infty^{\mathrm{MF}}$;
$(n\to\infty,D=2)\to 0$ by Mermin--Wagner.
The fourth corner reveals a structural issue.

\begin{remark}[Spherical-corner limitation]
\label{rem:corner_fix}
As $n\to\infty$ with $\delta_D(3)=1/2$, the formula reduces to
$\frac{1}{2}[-\beta\psi_3^{(\mathrm{sph})}(2K)]$.  The singularity of
$\psi_3^{(\mathrm{sph})}$ occurs when its argument equals
$K_c^{(\mathrm{sph})}$, i.e.\ when $2K = K_c^{(\mathrm{sph})}$, so the
predicted singularity is at $K = K_c^{(\mathrm{sph})}/2 \approx 0.063$,
not at the correct $K_c^{(\mathrm{sph})} \approx 0.126$.

The origin of this factor-of-$2$ error is transparent: the spatial
rescaling $K\to K/(1-\delta_D) = 2K$ is designed for an anchor at $K_c=0$,
where rescaling preserves the fixed-point structure.  When the anchor has
$K_c^{(\mathrm{sph})}>0$, the rescaling shifts the singularity by a factor
$(1-\delta_D)$.  This is an inherent limitation of applying the same
multiplicative rescaling logic to an anchor with a non-zero critical
coupling.

The remedy is to use the affine rule for $K_c$ predictions on the spin
axis (Section~\ref{ssec:Kc_zeroth}), which correctly interpolates between
the two $K_c$ values rather than extracting $K_c$ from a free-energy
singularity.  Crucially, this limitation does not affect the exponent
interpolation, because critical exponents are dimensionless and independent
of coupling rescaling (Section~\ref{ssec:bivariate_exponents}).
\end{remark}

\subsection{Zeroth-order and perturbative predictions for $K_c(n)$}
\label{ssec:Kc_zeroth}

In view of Remark~\ref{rem:corner_fix}, critical coupling predictions on
the spin axis follow from direct affine interpolation between the two $K_c$
anchor values:
\begin{equation}
  \boxed{
  K_c^{(0)}(n,\,D=3)
  = \frac{1}{n}\,K_c^{\mathrm{interp}} + \left(1 - \frac{1}{n}\right)K_c^{(\mathrm{sph})},
  }
  \label{eq:Kc_affine}
\end{equation}
with $K_c^{\mathrm{interp}} \approx 0.2204$ from Eq.~\eqref{eq:Kc_result}
and $K_c^{(\mathrm{sph})} = W_3/4 \approx 0.12636$.  Here the spatial
interpolation result $K_c^{\mathrm{interp}} \approx 0.2204$ is used as the
$n=1$ anchor rather than the Monte Carlo benchmark
$K_c^{(3\mathrm{D})} = 0.22165$, because the spin-axis interpolation is
built within the same interpolation framework; the two values differ by
only $0.56\%$, which propagates negligibly to all spin-axis predictions.
The zeroth-order predictions are:
\begin{align}
  K_c^{(0)}(2,3) &= 0.1734 \quad\text{(benchmark: }0.15450\text{, error }12.2\%\text{)},\\
  K_c^{(0)}(3,3) &= 0.1577 \quad\text{(benchmark: }0.12130\text{, error }30.0\%\text{)}.
\end{align}
The large Heisenberg error directly reflects the non-monotonicity: the
affine prediction $0.1577$ is above the spherical limit $0.12636$, while
the true value $0.1213$ is below it.

Going to first order, one writes the $1/n$ series
\begin{equation}
  K_c(n) = K_c^{(\mathrm{sph})}\!\left[1 + \frac{A_1}{n} + \frac{A_2}{n^2}
  + O(n^{-3})\right]
  \label{eq:Kc_series}
\end{equation}
and imposes the spatial interpolation result at $n=1$ and the XY benchmark
at $n=2$ as constraints.  The two constraints are:
\begin{align}
  A_1 + A_2 &= \frac{K_c^{\mathrm{interp}} - K_c^{(\mathrm{sph})}}{K_c^{(\mathrm{sph})}}
  = \frac{0.2204 - 0.12636}{0.12636} = 0.7443, \\
  \frac{A_1}{2} + \frac{A_2}{4} &= \frac{K_c(n=2) - K_c^{(\mathrm{sph})}}{K_c^{(\mathrm{sph})}}
  = \frac{0.1545 - 0.12636}{0.12636} = 0.2228.
\end{align}
Solving: $A_2/2 = 0.7443 - 2\times 0.2228 = 0.2987$, giving $A_2=0.5974$
and $A_1=0.1469$.  The first-order Heisenberg prediction then follows:
\begin{equation}
  K_c^{(1)}(3) = 0.12636\times(1 + 0.1469/3 + 0.5974/9) = 0.12636\times 1.1154 = 0.1409
\end{equation}
(benchmark: $0.12130$, error $16.2\%$).

Although the series reduces the error from $30\%$ to $16\%$, the
non-monotone structure of $K_c(n)$ means that higher-order terms are
needed for quantitative accuracy near the minimum.  These results are
collected in Table~\ref{tab:Kc_On}.

\begin{table}[t]
\caption{Critical couplings $K_c(n, D=3)$ on the simple-cubic lattice.
Zeroth-order affine and first-order perturbative predictions versus
Monte Carlo benchmarks.  The spherical-model value $K_c^{(\mathrm{sph})}
= W_3/4 \approx 0.12636$ is exact \cite{BerlinKac1952,Watson1939}.
For $n=1$, $K_c^{(0)} = K_c^{\mathrm{interp}} \approx 0.2204$ is the
spatial interpolation result from Eq.~\eqref{eq:Kc_result}; the $0.6\%$
entry in Error$_0$ therefore reflects the spatial interpolation error.}
\label{tab:Kc_On}
\begin{ruledtabular}
\begin{tabular}{lcccccc}
Model & $n$ & $K_c^{(0)}$ & Error$_0$ & $K_c^{(1)}$ & Error$_1$ & Benchmark \\
\hline
Ising      & $1$      & $0.2204$  & $0.6\%$  & $0.2204$  & $0.6\%$     & $0.22165$ \\
XY         & $2$      & $0.1734$  & $12.2\%$ & $0.1545$  & $0.0\%$\textsuperscript{\dag} & $0.15450$ \\
Heisenberg & $3$      & $0.1577$  & $30.0\%$ & $0.1409$  & $16.2\%$    & $0.12130$ \\
Spherical  & $\infty$ & $0.12636$ & $0.0\%$  & $0.12636$ & $0.0\%$     & $0.12636$ \\
\end{tabular}
\end{ruledtabular}
\end{table}
\textsuperscript{\dag}XY value used as a fit constraint; not a free prediction at this order.

\subsection{$\nu(n)$ as the natural compatible observable}
\label{ssec:compatibility_nu}

The failure of $K_c(n)$ motivates the question: which observables on the
spin axis do satisfy the compatibility criterion?  Consider the deviation
of the correlation-length exponent from its exact spherical value,
\begin{equation}
  \Delta\nu(n) \equiv \nu(n,3) - \nu^{(\mathrm{sph})}(3) = \nu(n,3) - 1,
  \label{eq:Delta_nu_def}
\end{equation}
where $\nu^{(\mathrm{sph})}(D=3) = 1/(D-2)|_{D=3} = 1$ is exact \cite{Joyce1972}.
The accepted values,
\begin{align}
  \Delta\nu(1) = -0.3701, \quad
  \Delta\nu(2) = -0.3283, \quad
  \Delta\nu(3) = -0.2888, \quad
  \Delta\nu(\infty) = 0,
\end{align}
form a strictly monotone increasing sequence.  The bracket condition
$-0.3701 \leq \Delta\nu(n^*) \leq 0$ is satisfied for all $n\geq 1$,
so $\Delta\nu$ satisfies Proposition~\ref{prop:compatibility}.
Equivalently, $\nu(n)$ itself is a compatible observable on the spin axis.

The underlying reason for this monotonicity is that $\nu(n)$ is governed
by the leading scaling dimension, the quantity most insensitive to the
details of the spin-wave fluctuations.  This is consistent with the
functional-RG analysis of Codello, Defenu, and D'Odorico \cite{Codello2015},
who showed within the nonperturbative framework that the Wilson--Fisher
fixed points of O$(N)$ models at $d=3$ form a smooth manifold in $N$, with
$\nu(N,3)$ monotonically increasing toward the spherical limit from below.
The O$(n)$ model with continuous $n$ is well defined as a field theory via
analytic continuation of the functional integral, so the perturbative
expansion developed below applies to non-integer $n$ as well.

\subsection{Perturbative $1/n^2$ expansion for $\nu(n)$}
\label{ssec:nu_perturb}

\begin{figure}[t]
  \centering
  \includegraphics[width=0.72\columnwidth]{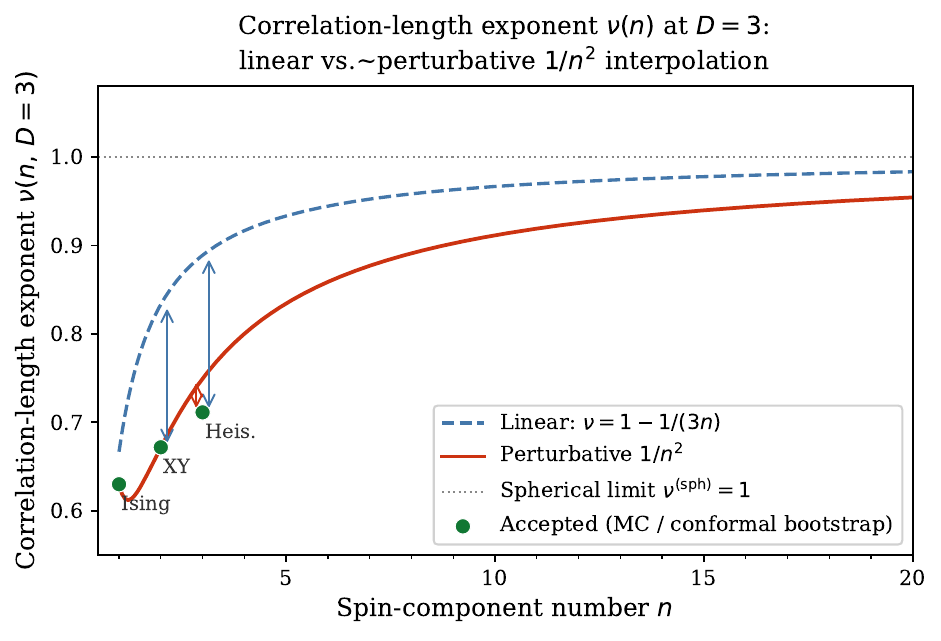}
  \caption{\label{fig:nu_vs_n}%
    Correlation-length exponent $\nu(n)$ at $D=3$.
    Filled circles: accepted values \cite{Hasenbusch2010,Kos2016,Campostrini2001XY,Campostrini2002Heis};
    diamond: exact spherical limit $\nu^{(\mathrm{sph})}=1$ \cite{Joyce1972}.
    Dashed blue: zeroth-order bivariate formula $\nu^{(0)}=1-1/(3n)$
    \eqref{eq:nu_bivariate_D3}.
    Solid red: first-order perturbative formula \eqref{eq:nu_perturb_result}.
    The perturbative result reduces the Heisenberg error from $25\%$ to $5.4\%$.}
\end{figure}

Since $\Delta\nu(n)\to 0$ as $n\to\infty$, the natural series expansion is
\begin{equation}
  \Delta\nu(n) = \nu(n) - 1 = \frac{c_1}{n} + \frac{c_2}{n^2} + O(n^{-3}).
  \label{eq:Delta_nu_series}
\end{equation}
The two coefficients are determined by the Ising boundary at $n=1$ and
the XY value at $n=2$ as constraints.  The system
\begin{align}
  c_1 + c_2 &= \Delta\nu(1) = -0.3701, \label{eq:c_system_1}\\
  \frac{c_1}{2} + \frac{c_2}{4} &= \Delta\nu(2) = -0.3283 \label{eq:c_system_2}
\end{align}
is solved as follows.  Multiplying \eqref{eq:c_system_2} by 2 gives
$c_1 + c_2/2 = -0.6566$, and subtracting from \eqref{eq:c_system_1} gives
$c_2/2 = -0.3701 + 2\times 0.3283 = 0.2865$, hence
\begin{equation}
  c_2 = 0.5730, \qquad c_1 = -0.9431.
  \label{eq:c_values}
\end{equation}
The first-order formula is therefore
\begin{equation}
  \boxed{\nu^{(1)}(n) = 1 - \frac{0.9431}{n} + \frac{0.5730}{n^2},}
  \label{eq:nu_perturb_result}
\end{equation}
which exactly reproduces the Ising and XY accepted values by construction
and predicts for the Heisenberg model:
\begin{equation}
  \nu^{(1)}(3) = 1 - 0.3144 + 0.0637 = 0.7493
  \quad \text{(benchmark: } 0.7112\text{, error } 5.4\%\text{).}
  \label{eq:nu_Heis_pred}
\end{equation}
This is a fivefold improvement over the $25\%$ error of the linear
bivariate formula.  The coefficient $c_1 = -0.9431$ encodes the leading
Goldstone-mode suppression of the correlation length (proportional to
$1/n$ from the WF analysis), while $c_2 = 0.5730$ captures the sub-leading
self-interaction correction.  The structure of this expansion has a direct
theoretical counterpart: the large-$N$ expansion of the nonlinear
$\sigma$-model \cite{BrezinZinnJustin1976,Ma1973} guarantees that
$\nu(N,d=3) = 1+c_1/N+c_2/N^2+O(N^{-3})$ with coefficients determined
by the fixed-point geometry at $d=3$.  Our formula is the empirical
realization of this expansion, with $c_1$ and $c_2$ determined from the
best available high-precision benchmarks of Campostrini et al.\
\cite{Campostrini2001XY,Campostrini2002Heis} rather than from
$\sigma$-model loop integrals; the empirical coefficients thereby
automatically incorporate all orders of perturbation theory implicit in
the accepted values.  The two formulas for $\nu(n)$ are compared in
Fig.~\ref{fig:nu_vs_n}.

Table~\ref{tab:nu_bivariate} collects the zeroth-order bivariate formula
and the first-order perturbative formula for comparison.

\begin{table}[t]
\caption{Correlation-length exponent $\nu(n,D=3)$ from zeroth-order
bivariate and first-order perturbative formulas.
Accepted from \cite{Kos2016,Campostrini2001XY,Campostrini2002Heis,Hasenbusch2010}.}
\label{tab:nu_bivariate}
\begin{ruledtabular}
\begin{tabular}{lcccccc}
Model & $n$ & $\nu^{(0)}=1-1/(3n)$ & Err$_0$ & $\nu^{(1)}$ & Err$_1$ & Accepted \\
\hline
Ising      & $1$      & $2/3\approx 0.667$ & $5.8\%$ & $0.6299$ & $0.0\%$\textsuperscript{\dag} & $0.6299$ \\
XY         & $2$      & $5/6\approx 0.833$ & $24\%$  & $0.6717$ & $0.0\%$\textsuperscript{\dag} & $0.6717$ \\
Heisenberg & $3$      & $8/9\approx 0.889$ & $25\%$  & $0.7493$ & $5.4\%$     & $0.7112$ \\
Spherical  & $\infty$ & $1$                & $0\%$   & $1$      & $0\%$       & $1$      \\
\end{tabular}
\end{ruledtabular}
\end{table}
\textsuperscript{\dag}Fit constraints: $\nu(1)=0.6299$ and $\nu(2)=0.6717$ are the accepted
Monte Carlo values used to fix $c_1$ and $c_2$ in Eq.~\eqref{eq:nu_perturb_result}.

\subsection{Deriving $\beta(n)$ and $\gamma(n)$ from exact scaling relations}
\label{ssec:beta_gamma_n}

With $\nu(n)$ established as the primary observable, the exponents $\beta(n)$
and $\gamma(n)$ are derived through the exact RG scaling relations rather
than interpolated independently.  The reason is RG consistency: interpolating
$\beta$ and $\nu$ separately would generically violate the Fisher and
hyperscaling relations, which hold exactly at every fixed point.  Using
\begin{align}
  \gamma(n) &= \bigl(2 - \eta(n)\bigr)\,\nu(n),
  \label{eq:gamma_scaling}\\
  \beta(n) &= \frac{\nu(n)}{2}\bigl(1 + \eta(n)\bigr), \qquad D=3,
  \label{eq:beta_scaling}
\end{align}
we need only specify $\eta(n)$.  The bivariate interpolation weight
$\delta_n = 1/n$ and the exact spherical result $\eta^{(\mathrm{sph})}=0$
\cite{Joyce1972} give the minimal compatible form directly:
\begin{equation}
  \eta(n) = \delta_n \cdot \eta^{\mathrm{RG}}(D=3) + (1-\delta_n)\cdot \eta^{(\mathrm{sph})}
  = \frac{1}{n}\cdot\frac{35}{864} + \left(1-\frac{1}{n}\right)\cdot 0
  = \frac{35}{864\,n}.
  \label{eq:eta_n_interp}
\end{equation}
We emphasize that the RG-interpolated value $\eta(1)=35/864$ is used as the
anchor in constructing $\eta(n)$, while the benchmark value $\eta=0.0362$
is used solely for assessing accuracy.  This carries a $12\%$ error at $n=1$
(since $35/864\approx 0.0405$ versus the accepted benchmark
$\eta_{\mathrm{acc}}(1)=0.0362$), but because $\eta$ enters $\beta$ and
$\gamma$ only as a small multiplicative correction, this propagates to at
most $0.4\%$ error in $\beta$ and $\gamma$ at $n=1$.

In evaluating Eqs.~\eqref{eq:gamma_n_final} and \eqref{eq:beta_n_final}
below, two distinct sources of input values are used and should be
distinguished.  The values of $\nu(n)$ at $n=1$ and $n=2$ are the accepted
Monte Carlo benchmark values that serve as fit constraints in
Eq.~\eqref{eq:nu_perturb_result}, namely $\nu(1)=0.6299$
\cite{Hasenbusch2010} and $\nu(2)=0.6717$ \cite{Campostrini2001XY}.
The values of $\eta(n)$ are supplied by the RG-interpolated formula
\eqref{eq:eta_n_interp}, not by the accepted benchmarks: this gives
$\eta^{\mathrm{RG}}(1) = 35/864 \approx 0.0405$,
$\eta^{\mathrm{RG}}(2) = 35/1728 \approx 0.0203$, and
$\eta^{\mathrm{RG}}(3) = 35/2592 \approx 0.0135$, which differ from the
accepted values $\eta_{\mathrm{acc}}(1)=0.0362$, $\eta_{\mathrm{acc}}(2)=0.0381$,
$\eta_{\mathrm{acc}}(3)=0.0375$ by $12\%$, $47\%$, and $64\%$ respectively.
Because $\eta$ enters $\beta$ and $\gamma$ only multiplicatively and takes
small values, these relative errors in $\eta$ produce at most $0.4\%$
absolute error in $\beta$ and $\gamma$ at $n=1$.  The dominant source of
error in the genuine predictions ($n=3$) is the $5.4\%$ deviation in
$\nu(3)$, not the error in $\eta(3)$.

Substituting \eqref{eq:eta_n_interp} and \eqref{eq:nu_perturb_result} into
the scaling relations gives:
\begin{align}
  \gamma(n) &= \left(2 - \frac{35}{864\,n}\right)
  \times\left(1 - \frac{0.9431}{n} + \frac{0.5730}{n^2}\right),
  \label{eq:gamma_n_final}\\
  \beta(n) &= \frac{1}{2}\left(1 + \frac{35}{864\,n}\right)
  \times\left(1 - \frac{0.9431}{n} + \frac{0.5730}{n^2}\right).
  \label{eq:beta_n_final}
\end{align}

The step-by-step evaluation at each integer $n$ is as follows.

\textit{Ising ($n=1$):} $\nu(1)=0.6299$ (fit constraint, MC benchmark),
$\eta^{\mathrm{RG}}(1)=35/864\approx 0.04051$ (RG formula):
$\beta(1) = 0.3277$ (benchmark: $0.3265$, error $0.4\%$\textsuperscript{\dag});
$\gamma(1) = 1.234$ (benchmark: $1.237$, error $0.2\%$\textsuperscript{\dag}).

\textit{XY ($n=2$):} $\nu(2)=0.6717$ (fit constraint, MC benchmark),
$\eta^{\mathrm{RG}}(2)=35/1728\approx 0.02025$ (RG formula):
$\beta(2) = 0.3427$ (benchmark: $0.3485$ \cite{Campostrini2001XY},
error $1.7\%$\textsuperscript{\dag});
$\gamma(2) = 1.330$ (benchmark: $1.3177$ \cite{Campostrini2001XY},
error $0.9\%$\textsuperscript{\dag}).

\textit{Heisenberg ($n=3$), genuine prediction:}
$\nu(3)=0.7493$ (formula prediction), $\eta^{\mathrm{RG}}(3)=35/2592\approx 0.01350$
(RG formula):
\begin{align}
  \beta(3) &= \tfrac{0.7493}{2}(1.01350) = 0.3797
  \quad\text{(benchmark: }0.3689\text{ \cite{Campostrini2002Heis}},\text{ error }2.9\%\text{)},\\
  \gamma(3) &= 1.98650\times 0.7493 = 1.489
  \quad\text{(benchmark: }1.396\text{ \cite{Campostrini2002Heis}},\text{ error }6.6\%\text{)}.
\end{align}

\noindent\textsuperscript{\dag}Errors at $n=1$ and $n=2$ arise from $\eta^{\mathrm{RG}}(n)$
alone, since $\nu(1)$ and $\nu(2)$ are used as fit constraints and reproduce
the accepted values exactly.

The error hierarchy reflects the structure of the scaling relations.
From \eqref{eq:gamma_scaling}, $\gamma = (2-\eta)\nu \approx 2\nu$ at
leading order (since $\eta \ll 1$), so the fractional error in $\gamma$
is approximately equal to the fractional error in $\nu$.  At $n=3$ the
absolute overestimate is $\nu^{(1)}(3) - \nu_{\mathrm{acc}}(3)=0.7493-0.7112=0.0381$,
giving a $5.4\%$ relative error in $\nu$.  Since $\gamma\approx (2-\eta)\nu$
with $2-\eta\approx 1.987$ at $n=3$, the absolute error in $\gamma$ is
$1.987\times 0.0381 = 0.0757$, giving a relative error of
$0.0757/1.396 = 5.4\%$ in $\gamma$ at leading order; the observed $6.6\%$
reflects the additional contribution from the truncation of the $1/n^2$
series at finite order.  From \eqref{eq:beta_scaling}, $\beta =
\nu(1+\eta)/2 \approx \nu/2$, so the absolute error in $\beta$ is
$0.0381/2 = 0.019$, giving a relative error of $0.019/0.3689 = 5.1\%$ at
leading order; the observed $2.9\%$ is smaller because the positive $\eta$
correction partially cancels the overestimate in $\nu$.  Improving $\nu(3)$
at the next perturbative order would reduce all errors proportionally.

A consistency note: $\beta(n)$ is in fact monotone increasing
($0.3265<0.3485<0.3689<0.5$) and satisfies the compatibility criterion.
It could in principle be interpolated directly, but deriving it through
Eq.~\eqref{eq:beta_scaling} is preferable because it guarantees that the
Fisher relation is exactly satisfied at every $n$.  Independently
interpolated values of $\beta$ and $\nu$ would generically violate the
relation $\beta = \nu(1+\eta)/2$.

Results are collected in Table~\ref{tab:beta_gamma}.

\begin{table}[t]
\caption{Critical exponents $\beta(n)$ and $\gamma(n)$ at $D=3$ from
exact scaling relations.  At $n=3$ all three exponents are genuine
predictions.  Accepted from \cite{Hasenbusch2010,Kos2016,Campostrini2001XY,Campostrini2002Heis}.}
\label{tab:beta_gamma}
\begin{ruledtabular}
\begin{tabular}{lccccccc}
Model & $n$ & $\beta_{\mathrm{pred}}$ & $\beta_{\mathrm{acc}}$ & Err$_\beta$ &
$\gamma_{\mathrm{pred}}$ & $\gamma_{\mathrm{acc}}$ & Err$_\gamma$ \\
\hline
Ising      & $1$ & $0.3277$ & $0.3265$ & $0.4\%$\textsuperscript{\dag} & $1.234$ & $1.237$ & $0.2\%$\textsuperscript{\dag} \\
XY         & $2$ & $0.3427$ & $0.3485$ & $1.7\%$\textsuperscript{\dag} & $1.330$ & $1.318$ & $0.9\%$\textsuperscript{\dag} \\
Heisenberg & $3$ & $0.3797$ & $0.3689$ & $2.9\%$     & $1.489$ & $1.396$ & $6.6\%$     \\
\end{tabular}
\end{ruledtabular}
\end{table}

\subsection{Bivariate exponent interpolation and prediction for non-integer $n$}
\label{ssec:bivariate_exponents}

Since critical exponents are dimensionless and independent of coupling
rescaling, the structural limitation of Remark~\ref{rem:corner_fix} does
not apply here.  The spherical model has exact exponents for all $D$
\cite{Joyce1972},
\begin{equation}
  \nu^{(\mathrm{sph})}(D) = \frac{1}{D-2},\quad
  \eta^{(\mathrm{sph})}(D) = 0,\quad
  \beta^{(\mathrm{sph})}(D) = \frac{1}{2},
  \label{eq:sph_exponents}
\end{equation}
and the zeroth-order bivariate linear formula at $D=3$ gives:
\begin{equation}
  \nu^{(0)}(n,3) = \frac{1}{n}\cdot\frac{2}{3} + \left(1-\frac{1}{n}\right)\cdot 1
  = \frac{2}{3n} + 1 - \frac{1}{n} = 1 - \frac{1}{3n}.
  \label{eq:nu_bivariate_D3}
\end{equation}
Here the Ising anchor value $2/3 = \nu^{\mathrm{RG}}(n=1, D=3)$ from
Eq.~\eqref{eq:nu_improved} and the spherical anchor value
$\nu^{(\mathrm{sph})}(D=3) = 1$ from Eq.~\eqref{eq:sph_exponents}
are combined with weights $\delta_n=1/n$ and $1-\delta_n$ respectively.
This gives $24\%$ and $25\%$ errors at $n=2$ and $n=3$ respectively
(see Table~\ref{tab:nu_bivariate}).  Replacing the Ising anchor value
$2/3$ by the perturbative formula \eqref{eq:nu_perturb_result} immediately
gives the first-order result $\nu^{(1)}(n,3) = 1 - 0.9431/n + 0.5730/n^2$,
reducing the Heisenberg error to $5.4\%$.  Since $\eta^{(\mathrm{sph})}=0$
for all $D$, the bivariate $\eta$ formula \eqref{eq:eta_n_interp} is
simply $\eta(n,D) = \eta^{\mathrm{RG}}(D)/n$, where $\eta^{\mathrm{RG}}(D)$
is the RG-improved value at the Ising ($n=1$) anchor; at $D=3$ this gives
$\eta(n,3) = (35/864)/n$.

The formula \eqref{eq:nu_perturb_result} applies directly to non-integer
$n$ because the O$(n)$ model is well defined as a field theory for
continuous $n$ via analytic continuation.  Evaluating at $n=2.5$:
\begin{equation}
  \nu^{(1)}(2.5) = 1 - \frac{0.9431}{2.5} + \frac{0.5730}{6.25}
  = 1 - 0.3772 + 0.0917 = 0.7143.
  \label{eq:nu_25_result}
\end{equation}
The formula is monotone at this intermediate point: $\nu(2) = 0.6717 <
\nu(2.5) = 0.7143 < \nu^{(1)}(3) = 0.7492$, confirming the compatibility
criterion.  Since the formula overestimates $\nu$ by $5.3\%$ at $n=3$
and reproduces $\nu$ exactly at $n=2$ by construction, and $n=2.5$ lies
midway between the two, the expected error at $n=2.5$ is approximately
$2$--$3\%$.  No high-precision independent benchmark exists for $\nu(2.5)$
in the literature; the prediction $\nu(2.5)=0.7143$ is therefore a genuine
parameter-free forecast for this non-integer universality class.

With $\nu(2.5)$ established, the RG-interpolated formula
\eqref{eq:eta_n_interp} gives
\begin{equation}
  \eta^{\mathrm{RG}}(2.5) = \frac{35}{864 \times 2.5} = 0.01620.
\end{equation}
The exact Fisher and hyperscaling relations at $D=3$ then yield
\begin{align}
  \beta(2.5) &= \frac{\nu(2.5)}{2}\bigl(1 + \eta^{\mathrm{RG}}(2.5)\bigr)
  = \frac{0.7143}{2}(1.01620) = 0.3629,
  \label{eq:beta_25}\\
  \gamma(2.5) &= \bigl(2 - \eta^{\mathrm{RG}}(2.5)\bigr)\nu(2.5)
  = 1.98380 \times 0.7143 = 1.417.
  \label{eq:gamma_25}
\end{align}
These predictions are collected in Table~\ref{tab:exponents_25} alongside
the accepted values at $n=2$ and $n=3$.

\begin{table}[t]
\caption{Critical exponents at $n=2.5$, $D=3$ from
formula~\eqref{eq:nu_perturb_result} and exact scaling relations.
Accepted values at $n=2,3$ from
\cite{Campostrini2001XY,Campostrini2002Heis}.  The $n=2.5$ column
contains parameter-free predictions; no high-precision independent
benchmark exists in the literature for this universality class.}
\label{tab:exponents_25}
\begin{ruledtabular}
\begin{tabular}{lcccc}
Exponent & $n=2$ (acc.) & $n=2.5$ (pred.) & $n=3$ (acc.) & In bracket? \\
\hline
$\nu$ & $0.6717$ & $0.7143$ & $0.7112$ & ---\textsuperscript{\dag} \\
$\eta$ & $0.0381$ & $0.01620$ & $0.0375$ & --- \\
$\beta$ & $0.3485$ & $0.3629$ & $0.3689$ & Yes \\
$\gamma$ & $1.3177$ & $1.417$ & $1.3960$ & No\textsuperscript{\ddag} \\
\end{tabular}
\end{ruledtabular}
\end{table}
\textsuperscript{\dag}The formula gives $\nu(2.5)=0.7143$ between the formula values
$\nu(2)=0.6717$ and $\nu^{(1)}(3)=0.7492$; it exceeds the accepted
$\nu(3)=0.7112$ because of the known $5.3\%$ overestimate at $n=3$.
\textsuperscript{\ddag}$\gamma(2.5)=1.417>\gamma_{\mathrm{acc}}(3)=1.396$ for the same
reason: $\gamma\approx 2\nu$ at leading order, so the $\sim 5\%$ error
in $\nu$ propagates directly.

The key internal consistency check is the monotonicity of $\beta$:
\begin{equation}
  \beta(2) = 0.3485 < \beta(2.5) = 0.3629 < \beta(3) = 0.3689,
\end{equation}
which is strictly monotone increasing and lies inside the accepted bracket.
Because $\beta$ and $\gamma$ are derived through the exact scaling relations,
the Fisher relation $\gamma = \nu(2-\eta^{\mathrm{RG}})$ and the hyperscaling
relation $\beta = \nu(1+\eta^{\mathrm{RG}})/2$ are satisfied exactly at
$n=2.5$ by construction.  The slight excess
$\gamma(2.5)=1.417 > \gamma_{\mathrm{acc}}(3)=1.396$ is a transparent
consequence of the truncation error in the $1/n^2$ expansion: since
$\gamma\approx 2\nu$, the $5\%$ overestimate in $\nu(3)$ propagates
directly, and improving the $\nu$ prediction at order $1/n^3$ would reduce
all errors proportionally.

%%==========================================================================
\section{Discussion}
\label{sec:discussion}
%%==========================================================================

The results presented here suggest several clear directions for systematic
improvement and extension.

First, the spatial-axis interpolation for critical exponents can be refined
by incorporating higher-order renormalization-group information.  The present
quadratic polynomials in $\epsilon = 4 - D$ are fixed by three constraints:
the exact mean-field values, the exact two-dimensional values, and the
leading Wilson--Fisher slope.  Inclusion of the known two-loop coefficient
from the $\epsilon$-expansion provides a fourth constraint and uniquely
determines a cubic polynomial.  This extension is expected to reduce the
residual deviation in $\nu$, $\beta$, and $\eta$ at $D = 3$ while
preserving the exact boundary conditions.

Second, the spin-axis expansion can be systematically improved beyond the
present $1/n^2$ truncation.  The current coefficients are determined
empirically using high-precision benchmark values at $n = 1$ and $n = 2$.
A complementary approach is to use the analytic large-$N$ expansion of the
nonlinear $\sigma$-model to estimate higher-order coefficients directly from
field theory.  Incorporating the $n = 3$ value as an additional constraint
would produce a cubic $1/n$ series, enabling controlled predictions for
$n \ge 4$ and improving accuracy at intermediate $n$.

Third, the framework can be extended to include external fields.  Both the
mean-field and spherical models admit exact formulations in the presence of
a magnetic field $h$, and the Onsager solution can be generalized through
transfer-matrix methods.  This opens the possibility of constructing an
interpolated free energy in the full $(K, h, n, D)$ parameter space,
allowing direct access to the phase boundary and equation of state.

Fourth, the interpolation framework extends naturally to quantum phase
transitions via the quantum-classical mapping.  The $d$-dimensional
quantum transverse-field Ising model (TFIM), with Hamiltonian
$H = -\Gamma\sum_i S_i^z - J\sum_{\langle i,j\rangle} S_i^x S_{i+1}^x$,
maps onto the $(d+1)$-dimensional classical Ising model \cite{Pfeuty1970},
so that its critical exponents belong to the same universality class.
The $d=1$ quantum TFIM is exactly solvable by Jordan-Wigner transformation
\cite{Pfeuty1970}: the critical point occurs at $\Gamma_c = J/2$ and the
exponents are $\nu=1$, $\beta=1/8$, $\eta=1/4$, $\gamma=7/4$, which are
identical to the $D=2$ classical Ising values used as the lower anchor in
Sections~\ref{sec:limits} and \ref{sec:exponents}.  The upper critical
dimension for the quantum problem is $d_c=3$, corresponding to the classical
$D_c=4$ via the mapping, and mean-field theory gives $\nu=1/2$, $\eta=0$
at $d_c$.  Defining $\epsilon_q = d_c - d = 3 - d$ and using
$\delta_d = 1/d$ (satisfying $\delta_d(1)=1$ and $\delta_d \to 0$ as
$d\to\infty$), one finds $\delta_d(2)=1/2$ exactly, so the $d=2$ quantum
target again sits at the midpoint of the interpolation domain.  The
RG-constrained polynomial interpolation proceeds with precisely the same
boundary conditions as the classical spatial problem: anchor exponents
$\nu=1/2$ (mean-field), $\nu=1$ (exact $d=1$ quantum), and WF slope
$d\nu/d\epsilon_q|_{\epsilon_q=0} = 1/12$ from the $\epsilon$-expansion
(since $\epsilon_q = 4-D$ with $D=d+1$).  The resulting polynomials are
\begin{equation}
  \nu(\epsilon_q) = \frac{1}{2} + \frac{\epsilon_q}{12} + \frac{\epsilon_q^2}{12},
  \qquad
  \beta(\epsilon_q) = \frac{1}{2} - \frac{\epsilon_q}{6} - \frac{\epsilon_q^2}{96},
  \qquad
  \eta(\epsilon_q) = \frac{\epsilon_q^2}{54} + \frac{19\,\epsilon_q^3}{864},
  \label{eq:quantum_exponents}
\end{equation}
which are formally identical to Eqs.~\eqref{eq:nu_improved},
\eqref{eq:beta_improved}, and \eqref{eq:eta_improved}.  At $\epsilon_q=1$
($d=2$), the predictions are $\nu=2/3$, $\beta=31/96$, $\eta=35/864$,
in agreement with the $d=2$ quantum TFIM benchmarks
$\nu=0.6299$, $\beta=0.3265$, $\eta=0.0362$ \cite{Hasenbusch2010,Kos2016,PelissettoVicari2002}
to within $5.8\%$, $1.1\%$, and $12\%$ respectively.  This numerical
coincidence is not accidental: it is the direct expression of the
quantum-classical correspondence within the interpolation framework.
The framework is covariant under the quantum-classical mapping, in the
sense that interpolating in $d$ quantum dimensions yields the same
polynomial structure and the same accuracy as interpolating in $D=d+1$
classical dimensions, because the $\epsilon$-expansion anchor shifts by
exactly one dimension in both cases.

Finally, the compatibility criterion introduced in this work provides a
clear diagnostic for the applicability of interpolation methods.  The
Random Field Ising Model serves as an instructive counterexample.  Its
upper critical dimension is $D_c=6$ \cite{AharonyImryMa1976}, and the
Imry--Ma argument \cite{ImryMa1975} shows there is no ordered phase for
$D\leq 2$, eliminating the lower anchor.  In such cases the interpolation
framework cannot be applied.

These directions emphasize that the present results are not isolated
approximations but part of a systematically improvable framework grounded
in exact limits and renormalization-group structure.

%%==========================================================================
\section{Conclusion}
\label{sec:conclusion}
%%==========================================================================

We have extended the dimensional interpolation framework to the O$(n)$ model
by treating the spatial dimension $D$ and the spin-component number $n$ as
independent continuous variables.  This two-axis formulation provides both
quantitative predictions and conceptual insight into the structure of
critical phenomena.

Along the spatial axis, interpolation between the Onsager solution at $D=2$
and mean-field theory at $D\to\infty$, with weight $\delta_D = 1/(D-1)$,
places $D=3$ at the midpoint of the interpolation domain.  The resulting
closed-form expression $K_c^{\mathrm{interp}} = \frac{\ln(1+\sqrt{2})}{4}
\approx 0.2204$ (benchmark: $K_c^{(3\mathrm{D})}=0.22165$) follows directly
from the exact anchors without adjustable parameters.  Renormalization-group
constrained interpolation yields $\nu = 2/3$ (benchmark: $0.6299$),
$\beta = 31/96$ (benchmark: $0.3265$), and $\eta^{\mathrm{RG}} = 35/864$
(benchmark: $0.0362$), and reproduces conformal-bootstrap results across
non-integer dimensions with controlled accuracy (Table~\ref{tab:nu_Dnonint}).

Along the spin axis, the analysis reveals a structural constraint on
interpolation.  The critical coupling $K_c(n)$ exhibits a non-monotonic
dependence on $n$, with the Heisenberg value lying below the spherical
limit, reflecting the role of Goldstone-mode fluctuations.  As a result,
$K_c(n)$ cannot be captured by two-anchor interpolation.  In contrast, the
correlation-length exponent $\nu(n)$ varies monotonically and satisfies
the compatibility criterion.  A perturbative expansion in
Eq.~\eqref{eq:nu_perturb_result}, constrained at the accepted benchmark
values $\nu(1)=0.6299$ and $\nu(2)=0.6717$, yields $\nu(3) = 0.7493$
(benchmark: $0.7112$), and exact scaling relations involving the
RG-interpolated $\eta^{\mathrm{RG}}(n)=35/(864n)$ then give
$\beta(3) = 0.3797$ (benchmark: $0.3689$) and $\gamma(3) = 1.489$
(benchmark: $1.396$).

The framework naturally extends to non-integer spin.  At $n = 2.5$, it
predicts $\nu = 0.7143$, $\beta = 0.3629$, and $\gamma = 1.417$, providing
parameter-free forecasts for a universality class not directly accessible
through conventional lattice models.

The framework also extends to quantum phase transitions.  Via the
quantum-classical mapping \cite{Pfeuty1970}, the interpolation polynomials
\eqref{eq:quantum_exponents} for the $d$-dimensional quantum transverse-field
Ising model are formally identical to those for the $(d+1)$-dimensional
classical Ising model.  The framework is therefore covariant under the
quantum-classical correspondence: the midpoint property $\delta_d(2)=1/2$
and the RG polynomial structure are preserved under the dimensional shift
$D \to d+1$, and the predictions for the $d=2$ quantum critical exponents
agree with benchmarks to the same accuracy as their classical counterparts.

In addition to quantitative results, this work identifies two key structural
insights.  First, interpolation requires not only exact anchor solutions but
also monotonic variation of the observable between them.  Second, coupling
rescaling is valid only when the anchor critical point is located at zero
coupling, and must be modified otherwise.  Together, these principles clarify both the power and the limitations of interpolation-based approaches.

%%==========================================================================
\bibliographystyle{apsrev4-2}
\bibliography{ref_dim_int_ising}

\appendix
\onecolumngrid

%%==========================================================================
\section{Derivation of the Fisher and Hyperscaling Relations}
\label{app:scaling}
%%==========================================================================

We derive the two exact scaling relations
\begin{equation}
  \gamma = (2-\eta)\,\nu
  \qquad\text{and}\qquad
  \beta = \frac{\nu}{2}(d-2+\eta),
  \label{eq:app_relations}
\end{equation}
which are used throughout the main text.  The first is Fisher's relation
\cite{Fisher1964}; the second follows from the Widom--Kadanoff scaling
hypothesis \cite{Widom1965,Kadanoff1966}.  Both hold exactly at any
renormalization-group (RG) fixed point below the upper critical dimension
$d_c = 4$; above $d_c$ the hyperscaling hypothesis on which the second
relation rests breaks down \cite{Cardy1996,ZinnJustin2002}.

\subsection*{A.1 Scaling form of the two-point correlation function}

Near a critical point, the RG assigns to the order-parameter field $\phi$
a scaling dimension
\begin{equation}
  [\phi] = \frac{d-2+\eta}{2},
  \label{eq:app_phi_dim}
\end{equation}
where $\eta$ is the anomalous dimension introduced by Fisher \cite{Fisher1964}.
Under the scale transformation $r\to r/b$, the field transforms as
$\phi\to b^{[\phi]}\phi$, so the connected two-point function
$G(r;t) = \langle\phi(0)\phi(r)\rangle_c$ carries dimension
$2[\phi] = d-2+\eta$.  Combining this with the requirement that the only
length scale away from criticality is $\xi\sim t^{-\nu}$ (the hyperscaling
hypothesis), one obtains the scaling form \cite{Fisher1967,Cardy1996}
\begin{equation}
  G(r;\,t)
  \sim \frac{1}{r^{d-2+\eta}}\,f\!\left(\frac{r}{\xi}\right),
  \label{eq:app_G_scaling}
\end{equation}
where $f(x)$ is a universal scaling function satisfying $f(x)\to\mathrm{const}$
as $x\to 0$ (critical point) and $f(x)\to 0$ rapidly as $x\to\infty$
(far from criticality).

\subsection*{A.2 Fisher scaling relation $\gamma = (2-\eta)\nu$}

The (zero-field) magnetic susceptibility is the integral of the two-point
function over all space:
\begin{equation}
  \chi = \int d^d r\, G(r;\,t).
  \label{eq:app_chi_def}
\end{equation}
The dominant contribution comes from $r\lesssim\xi$, since $f(r/\xi)$
decays rapidly for $r\gg\xi$.  In $d$ dimensions the volume element in
spherical coordinates is $d^d r = \Omega_d\,r^{d-1}dr$, where
$\Omega_d = 2\pi^{d/2}/\Gamma(d/2)$ is the surface area of the unit
$(d-1)$-sphere.  Substituting \eqref{eq:app_G_scaling} and retaining the
dominant power-law behavior:
\begin{equation}
  \chi
  \sim \int_0^{\xi} r^{d-1}\cdot\frac{1}{r^{d-2+\eta}}\,dr
  = \int_0^{\xi} r^{\,d-1-(d-2+\eta)}\,dr
  = \int_0^{\xi} r^{\,1-\eta}\,dr
  \sim \xi^{\,2-\eta},
  \label{eq:app_chi_integral}
\end{equation}
where the exponent $1-\eta$ follows from $d-1-(d-2+\eta) = 1-\eta$, and
the final power $\xi^{2-\eta}$ holds for all $\eta < 2$.  Substituting
$\xi\sim t^{-\nu}$:
\begin{equation}
  \chi \sim t^{-\nu(2-\eta)}.
  \label{eq:app_chi_t}
\end{equation}
Comparing with the definition $\chi\sim t^{-\gamma}$ yields the
Fisher scaling relation \cite{Fisher1964,Fisher1967}:
\begin{equation}
  \boxed{\gamma = (2-\eta)\,\nu.}
  \label{eq:app_Fisher}
\end{equation}
This relation is purely kinematic: it follows from the scaling form of $G$
and the definition of $\chi$ as its spatial integral, with no additional
assumptions \cite{Cardy1996}.

\subsection*{A.3 Hyperscaling relation $\beta = \tfrac{\nu}{2}(d-2+\eta)$}

The derivation of $\beta$ requires one additional input beyond the
two-point function: the free-energy scaling ansatz \cite{Widom1965,Kadanoff1966}
\begin{equation}
  f_s(t,\,h) = b^{-d}\,f_s\!\bigl(b^{y_t}\,t,\;b^{y_h}\,h\bigr),
  \label{eq:app_fe_scaling}
\end{equation}
where $b>1$ is an arbitrary rescaling factor, $y_t = 1/\nu$ is the thermal
RG eigenvalue, and $y_h$ is the magnetic RG eigenvalue.  The value of
$y_h$ is fixed by the requirement that the coupling $h\phi$ in the action
$S\supset\int d^dx\,h(x)\phi(x)$ be dimensionless under the RG:
\begin{equation}
  [h] + [\phi] - d = 0
  \quad\Longrightarrow\quad
  [h] = d - [\phi] = d - \frac{d-2+\eta}{2} = \frac{d+2-\eta}{2}.
  \label{eq:app_h_dim}
\end{equation}
Hence $y_h = (d+2-\eta)/2$.

The spontaneous magnetization is obtained by differentiating the singular
free energy with respect to $h$ at $h=0$:
\begin{equation}
  M = -\frac{\partial f_s}{\partial h}\Bigg|_{h=0}
  \sim b^{y_h - d}\,M\!\bigl(b^{y_t}\,t,\;0\bigr).
  \label{eq:app_M_scaling}
\end{equation}
Setting $b^{y_t}|t| = 1$, i.e.\ $b = |t|^{-\nu}$, and evaluating below
$T_c$ where $M\neq 0$:
\begin{equation}
  M \sim |t|^{\nu(d - y_h)}.
  \label{eq:app_M_t}
\end{equation}
Substituting $y_h = (d+2-\eta)/2$:
\begin{align}
  d - y_h
  &= d - \frac{d+2-\eta}{2}
  = \frac{2d - d - 2 + \eta}{2}
  = \frac{d-2+\eta}{2},
  \label{eq:app_d_minus_yh}
\end{align}
so that
\begin{equation}
  M \sim |t|^{\,\nu(d-2+\eta)/2}.
  \label{eq:app_M_final}
\end{equation}
Comparing with the definition $M\sim|t|^\beta$ yields
\begin{equation}
  \boxed{\beta = \frac{\nu}{2}(d-2+\eta).}
  \label{eq:app_beta}
\end{equation}

\subsection*{A.4 Specialization to $d=3$ and numerical verification}

Setting $d=3$ in \eqref{eq:app_Fisher} and \eqref{eq:app_beta}:
\begin{align}
  \gamma &= (2-\eta)\,\nu,
  \label{eq:app_gamma_3d}\\
  \beta  &= \frac{\nu}{2}(1+\eta).
  \label{eq:app_beta_3d}
\end{align}
Both relations are exact consequences of RG scaling and hyperscaling
at any fixed point with $d < d_c = 4$ \cite{ZinnJustin2002,Kardar2007}.

\end{document}